# Dirac magnons in honeycomb ferromagnets

Sergey S. Pershoguba[1], Saikat Banerjee[1,2,3], J.C. Lashley[2], Jihwey Park[4],

Hans Ågren[3], Gabriel Aeppli[4,5,6], and Alexander V. Balatsky[1,2,7]

[1] Nordita, Center for Quantum Materials, Royal Institute of Technology, and Stockholm University, Roslagstullsbacken 23, S-106 91 Stockholm, Sweden
[2] Institute for Materials Science, Los Alamos National Laboratory, Los Alamos, New Mexico 87545, USA
[3] Division of Theoretical Chemistry and Biology, Royal Institute of Technology, SE-10691 Stockholm, Sweden
[4] Paul Scherrer Institute, CH-5232 Villigen PSI, Switzerland
[5] Laboratory for Solid State Physics, ETH Zurich, Zurich, CH-8093, Switzerland,
[6] Institut de Physique, EPF Lausanne, Lausanne, CH-1015, Switzerland
[7] ETH Institute for Theoretical Studies, ETH Zurich, CH-8092 Zurich, Switzerland

**The discovery of the Dirac electron dispersion in graphene [1] led to the question of the Dirac cone stability with respect to interactions. Coulomb interactions between electrons were shown to induce a logarithmic renormalization of the Dirac dispersion. With a rapid expansion of the list of compounds and quasiparticle bands with linear band touching [2], the concept of bosonic Dirac materials has emerged. We consider a specific case of ferromagnets consisting of the Van der Waals-bonded stacks of honeycomb layers, e.g chromium trihalides $CrX_3$ (X = F, Cl, Br and I), that display two spin wave modes with energy dispersion similar to that for the electrons in graphene. At the single particle level, these materials resemble their fermionic counterparts. However, how different particle statistics and interactions affect the stability of Dirac cones has yet to be determined. To address the role of interacting Dirac magnons, we expand the theory of ferromagnets beyond the standard Dyson theory [3, 4] to a case of non-Bravais honeycomb layers. We demonstrate that magnon-magnon interactions lead to a significant momentum-dependent renormalization of the bare band structure in addition to strongly momentum-dependent magnon lifetimes. We show that our theory qualitatively accounts for hitherto unexplained anomalies in a nearly half century old magnetic neutron scattering data for $CrBr_3$ [5, 6]. We also show that honeycomb ferromagnets display dispersive surface and edge states, unlike their electronic analogs.**

## I. Introduction

The observation of fermionic quasiparticles with Dirac dispersion was a key finding for graphene [7]. Since then, the list of materials exhibiting the Dirac and Weyl energy spectra for Fermions has been extended further. Materials hosting bosonic excitations with the Dirac cones have opened a new stage in investigation of the Dirac materials such as photonic crystals [8, 9], plasmonic systems [10], honeycomb arrays of superconducting grains [11] and magnets [12]. Magnets were studied in the past, especially in the form of transition metal trihalides $TMX_3$ (where X = F, Cl, Br, I and TM=Cr), which consist of weakly coupled honeycomb ferromagnetic planes. These materials have a potential as spin polarizing elements and exhibit a strong Kerr and Faraday effects.

Early spin-wave analysis revealed a Dirac crossing point in the dispersion in the honeycomb layers containing two magnetic atoms per unit cell. Neutron scattering measurements revealed hitherto unexplained anomalies [5, 6] in the boson (spin-wave) self-energies near the Dirac points.

The single-particle properties for both bosonic and fermionic Dirac materials derive from the tight-binding model on the honeycomb lattice and are, thus, identical. At the level of quantum statistics, however, there is a difference between Dirac fermions and bosons. For fermions, the excitations occur near the chemical potential, and one can focus on the low-energy Dirac cones shown in Fig. 1(a). In contrast, bosons are not subject to the Pauli exclusion principle. They can explore the entire momentum space, and excitations near zero energy dominate at low temperatures.

The importance of many-body effects was appreciated immediately for fermionic Dirac materials [13]. In particular, the Coulomb interaction between electrons, see Fig. 1(c), leads to a logarithmic renormalization of the Dirac cone velocity. This renormalization was verified by observing an anomalous dependence of the cyclotron frequency on the carrier concentration [14].

Similar analysis for the interacting bosons and their effects on the Dirac node dispersion has not been done systematically to our knowledge. Since the early paper by F. Bloch [15], the physics of magnons has remained a subject of an active experimental and theoretical research [16, 17, 18]. Despite an immense number of works, the important case of ferromagnets with non-Bravais lattices has received a relatively little attention. In the two milestone papers by Dyson [3, 4], magnon thermodynamics is discussed only for crystals with Bravais lattices. Therefore, such theories cannot be directly applied to the honeycomb lattice, a prominent example of the non-Bravais bipartite lattices.

As mentioned we focus on the Dirac bosons, as realized by Cr trihalides with ferromagnetic honeycomb lattices. We start with a 2D honeycomb ferromagnet, where the bosons are spin-waves (magnons) that form Dirac nodes, Fig. 1. We calculate the lowest order self-energy diagrams shown in Fig. 1(d). The Hartree self-energy gives a uniform renormalization of the energy bands consistent with the theory of Bloch [19]. We then consider both the real and imaginary parts of the self-energy, which give the energy renormalization and decay rate of magnon excitations. We find that interactions induce strong temperature- and momentum-dependent renormalization of the magnon bands near a Dirac node. Our results allow to explain the outstanding puzzle of spin-wave anomalies seen in $CrB_3$[5, 6].

We also discuss the role of topology and qualitative difference between bosons and fermions for the formation of the surface states. Bosons do not have a Pauli constraint and several bosons can occupy the same state. This has important consequences for single particle properties such as surface states, and for many-body renormalization of the Dirac dispersion. Dispersive states can exist on the surface of a ferromagnetic "graphite", while they do not for fermions [20, 21, 22].

Our result are applicable to the rapidly growing class of Van der Waals ferromagnets with non-Bravais lattices. We note recent experimental works by Gong et al. [23] and Huang et al. [24] that demonstrate the robustness of honeycomb ferromagnetic phase in a single layer materials including $CrI_3$. Recent theoretical works [25, 26] with the focus on antiferromagnets are also relevant for our discussion.

The outline of the paper is as follows. After Introduction in Section I, we present the free spin-wave theory in Section II. In Section III, we present the main result of the paper – the evaluation of the self-energies and their effect on magnon spectrum. In Section IV, we discuss properties of topological surface states and conclude in Section V.

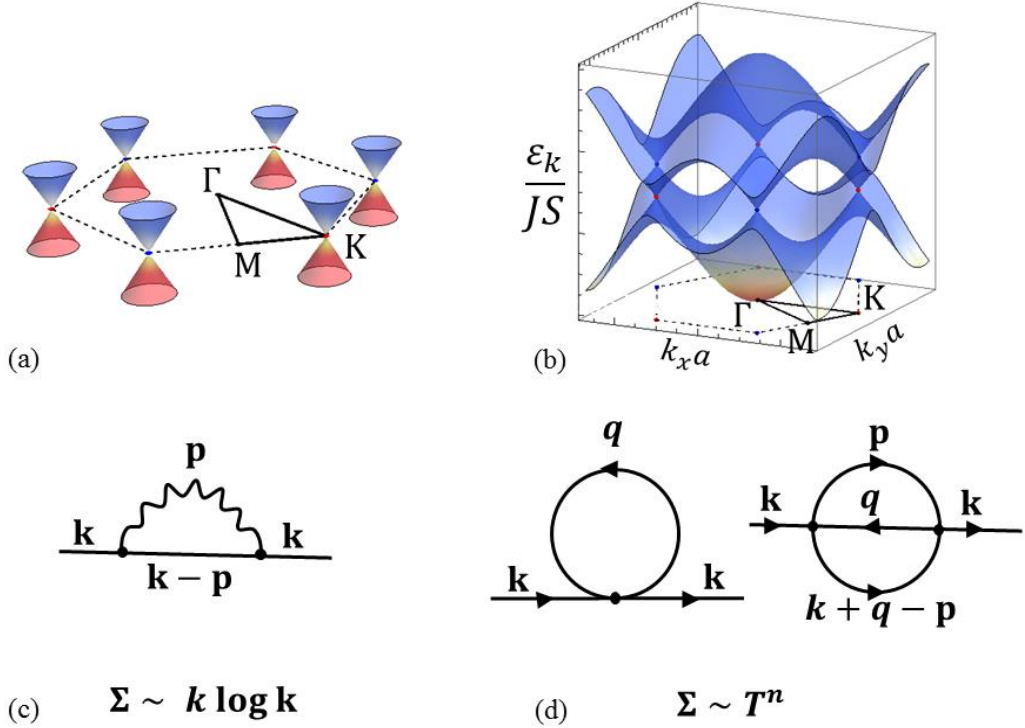

FIG 1: Comparison of (left panels) Dirac fermions vs (right panels) bosons. (a) The properties of the Dirac fermion materials are determined by the states at the chemical potential. So, approximating the band structure by the Dirac cones zone is sufficient. (b) In contrast, since bosons can freely occupy any states within the Brillouin zone, information about the entire band structure is necessary. In the top panels, the color scheme "red-to-blue" is an artistic representation of the particle occupation number from ``strong-to-weak''. Temperature can efficiently excite the bosons in the down band in the vicinity of $\Gamma$-point. (c) The relevant self-energy diagram that gives the logarithmic correction to the Dirac fermion velocity. (d) The relevant self-energy diagrams giving a temperature-dependent renormalization and decay of the Dirac bosons.

II. Model.

We begin with a Heisenberg model on a 2D honeycomb lattice with two sites per unit cell commonly denoted as A and B. The Heisenberg Hamiltonian of the model $\mathcal{H} = -J\sum_{\langle i,j \rangle}(\mathbf{S}_i \cdot \mathbf{S}_j)$ describes the nearest-neighbor coupling between spins. For convenience, we choose energy units for the coupling $J$, whereas the magnitude of spin S is dimensionless. Positive coupling $J > 0$ implies a ferromagnetic ground state at low temperatures ($J \sim T_c$ = 32.5 K, $S = 3/2$ for $CrBr_3$). We choose to study the magnetic excitations, the Dirac magnons, above the preexisting ferromagnetic ground state on the honeycomb lattice. For the moment we ignore the interlayer coupling in $CrBr_3$ that will be discussed at the end of the paper.

We follow the standard practice and bosonize the Heisenberg Hamiltonian using the Holstein-Primakoff transformation. Truncated to zeroth order, these transformations relate the spin-operators to the magnon creation/annihilation operators as $S^x + iS^y = \sqrt{2S}\, a$, $S^x + iS^y = \sqrt{2S}\, a^\dagger$, and $S_i^z = S - a_i^\dagger a_i$. We thus obtain the free bosonic Hamiltonian

$$\mathcal{H} = \sum_{\mathbf{k}} \Psi_{\mathbf{k}}^\dagger H_0(\mathbf{k}) \Psi_{\mathbf{k}}, \qquad H_0(\mathbf{k}) = JS \begin{pmatrix} 3 & -\gamma_{\mathbf{k}} \\ -\gamma_{\mathbf{k}}^* & 3 \end{pmatrix}. \qquad (1)$$

Here, the two-by-two single-particle Hamiltonian $H_0$ acts upon the spinor $\Psi_k = (a_k, b_k)^T$ with components corresponding to the two sublattices. The off-diagonal element $\gamma_{\mathbf{k}} = \sum_j e^{i\mathbf{k}\mathbf{r}_j}$ is ubiquitous for the honeycomb lattice and corresponds to coupling along the three nearest-neighbor in-plane bond vectors $\mathbf{r}_j$. The energy spectrum of the single particle consists of two branches: the acoustic "down" and optical "up" branch with the dispersion $\varepsilon_k^{u,d} = JS(3 \pm |\gamma_k|)$ and wave functions $\Psi_{\mathbf{k}}^{u,d} = \left( e^{i\phi_{\mathbf{k}}/2}, s\, e^{-i\phi_{\mathbf{k}}/2} \right)/\sqrt{2}$, where $s = -1$ ($s = +1$) for the u (d) state and the momentum-dependent phase $\phi_{\mathbf{k}} = \mathrm{Arg}\,\gamma_{\mathbf{k}}$ is introduced. The magnon energy dispersion is plotted in Fig. 1(b). The down branch touches zero energy quadratically in the center of the Brillouin zone near the Γ point as $\varepsilon_{\mathbf{k}}^d = 3JSk^2/4$. The gapless magnons at the Γ point are protected by the Goldstone theorem. The magnon branches have symmetry-protected Dirac crossings at the K and K′ points of the Brillouin zone characteristic of the honeycomb lattice. At the single-particle level, the magnon dispersion law is identical to the energy spectrum of electrons in graphene.

### III. Results-bulk states

The interaction vertex is obtained from the next-order terms following from the Holstein-Primakoff transformation (see Supplemental Material for details)

$$\mathcal{V} = \frac{J}{4N} \sum_{\{\mathbf{k}\}} \gamma_{\mathbf{k}_2}^* a_{\mathbf{k}_1}^\dagger b_{\mathbf{k}_2}^\dagger a_{\mathbf{k}_3} a_{\mathbf{k}_4} + \gamma_{\mathbf{k}_4} a_{\mathbf{k}_1}^\dagger a_{\mathbf{k}_2}^\dagger a_{\mathbf{k}_3} b_{\mathbf{k}_4} \\
+ \gamma_{\mathbf{k}_2} b_{\mathbf{k}_1}^\dagger a_{\mathbf{k}_2}^\dagger b_{\mathbf{k}_3} b_{\mathbf{k}_4} + \gamma_{\mathbf{k}_4}^* b_{\mathbf{k}_1}^\dagger b_{\mathbf{k}_2}^\dagger b_{\mathbf{k}_3} a_{\mathbf{k}_4} - 4\gamma_{\mathbf{k}_4 - \mathbf{k}_2} a_{\mathbf{k}_1}^\dagger b_{\mathbf{k}_2}^\dagger a_{\mathbf{k}_3} b_{\mathbf{k}_4}, \qquad (2)$$

where the momentum before and after scattering is conserved $\mathbf{k}_1 + \mathbf{k}_2 = \mathbf{k}_3 + \mathbf{k}_4$. Note that the same function $\gamma_{\mathbf{k}}$ defined above for the single-particle Hamiltonian (1) also occurs in the interaction term (2).

We analyze the effect of interactions to first order and evaluate the Hartree diagram in Fig. 1(d). For simplicity, consider the case of low temperature T<<J, where only the low-energy down-magnons with momenta close to $\mathbf{q} = 0$ are excited. The Hartree term corresponds to contracting a pair of boson operators in Eq. (2) and replacing them with the thermodynamic boson occupation number of the magnons in the lower band $\langle d_{\mathbf{q}}^\dagger d_{\mathbf{q}} \rangle = f(\varepsilon_{\mathbf{q}}^d) = \left[ \exp(\varepsilon_{\mathbf{q}}^d / T) - 1 \right]^{-1}$. We further perform a low-temperature expansion and rewrite the self-energy

$$\Sigma_T^{(1)}(\mathbf{k}) = -\alpha_1 T^2 H_0(\mathbf{k}) \quad , \quad (3)$$

where $\alpha_1 = \dfrac{\pi}{24\sqrt{3}J^2 S^3}$. The matrix elements of the two-by-two Hamiltonian are renormalized by the same temperature-dependent ratio $\alpha_1 T^2$. This is a consequence of a delicate balance between the bare spectrum and interaction term, both containing the honeycomb function $\gamma_\mathbf{q}$. So, the energy of the renormalized bands becomes $(1-\alpha_1 T^2)\varepsilon_\mathbf{k}^{u,d}$. The magnon-magnon interaction leads to an overall temperature-dependent bandwidth renormalization. This is consistent with the observation by Bloch [19], who discussed a similar effect for a single-band case on a cubic lattice.

Next, we analyze the effect of interactions to second order. We consider the "rainbow" diagram in Fig. 1(d)

$$\Sigma^{(2)}(\omega,\mathbf{k}) = A^2 \int \frac{d^2q\, d^2p}{(2\pi)^4} \frac{|V_{\mathbf{k},\mathbf{q};\mathbf{p}}|^2 f(\varepsilon_\mathbf{q})}{\omega + \varepsilon_\mathbf{q} - \varepsilon_\mathbf{p} - \varepsilon_{\mathbf{k}+\mathbf{q}-\mathbf{p}} + i\delta} \quad , \quad (4)$$

where $V_{\mathbf{k},\mathbf{q};\mathbf{p}}$ is the scattering matrix element, and $A = 3\sqrt{3}a^2/2$ is the area of the honeycomb lattice unit cell. Let us comment on the labeling of the momenta in the scattering process: two original magnons with momenta $\mathbf{k},\mathbf{q}$ are scattered into the two magnons with momenta $\mathbf{p}$ and $\mathbf{k}+\mathbf{q}-\mathbf{p}$. The momentum of the thermally excited magnons is $\mathbf{q}$. Note that the numerator of the diagram contains the Bose occupation number of the thermal magnons $f(\varepsilon_\mathbf{q})$, which controls the temperature dependence of the self-energy. In the limit of small temperatures, there is a simplification: only the thermal magnons from the down band in the vicinity of the $\Gamma$ point are excited as is illustrated in Fig. 1(b). The low temperature expansion allows us to integrate out the thermal magnons $\mathbf{q}$ analytically and reduce Eq. (4) to

$$\Sigma_T^{(2)}(\omega,\mathbf{k}) = \alpha_2 T^2 A \int \frac{d^2 p}{(2\pi)^2} \frac{|v_{\mathbf{k};\mathbf{p}}|^2}{\omega - \varepsilon_{\mathbf{p}} - \varepsilon_{\mathbf{k}-\mathbf{p}} + i\delta}, \qquad (5)$$

where $\alpha_2 = \dfrac{\pi}{6\sqrt{3}J^2 S^2}$, and $v_{\mathbf{k};\mathbf{p}}$ is an effective matrix element at low temperature. The remaining sum in Eq. (5) is now carried out numerically and the result is plotted in Fig. 2 for momentum $\mathbf{k}$ along a closed path $\Gamma \to K \to M \to \Gamma$ in momentum space. Note that the magnitude of the diagram is set by the overall temperature-dependent prefactor in Eq. (5). We show the self-energy separately for the up and down magnon branches in the left and right columns of Fig. 2, correspondingly. Generally, we find that the matrix element in the numerator is a well-behaved nonsingular function in momentum space that only slightly modifies the pattern of the self-energy. In contrast, the energy denominator in Eq. (5) is the major factor determining the self-energy (Fig.2) as a function of momentum. The real and imaginary parts of the self-energy have physical significance as the scattering rate and energy renormalization of the magnon excitations, respectively.

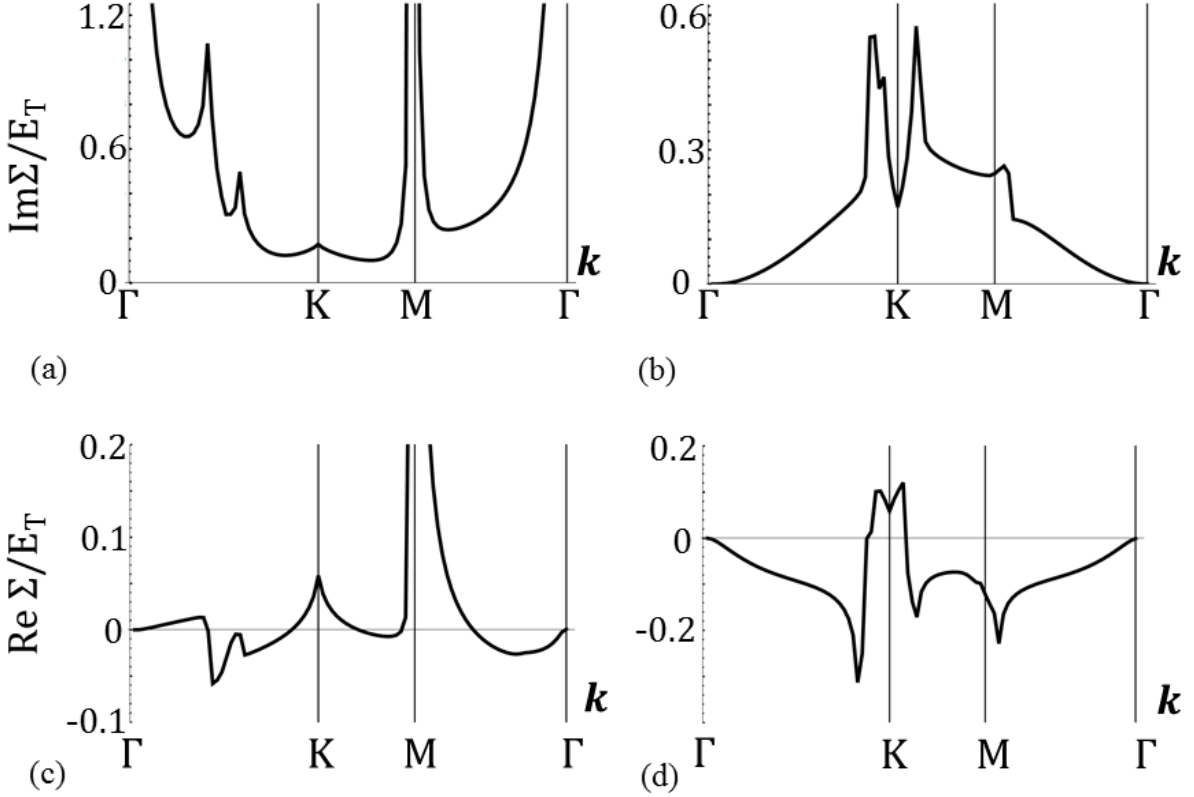

FIG 2: Normalized decay rate of the (a) up-band and (b) down-band. A sequence of van-Hove like peaks is a unique signature of the Dirac dispersion. Normalized magnon dispersion renormalization for (c) up-band and (d) down-band. The vertical axes are plotted with respect to temperature- dependent energy unit $E_T = T^2/JS^3$; therefore all contributions scale as $T^2$ with temperature T.

First, let us comment on the scattering rate, i.e. the imaginary part of the self-energy, shown in the top row of Fig. 2. The scattering process is on-shell, i.e. the energy of the magnons is conserved. Due to the multiband nature of the Hamiltonian, there are multiple scattering channels contributing to each decay process. For example, the scattering rate for the up band shown in Fig. 2(a) is derived from the two scattering channels $u_{\mathbf{k}} \to u_{\mathbf{p}} + d_{\mathbf{k}-\mathbf{p}}$ and $u_{\mathbf{k}} \to d_{\mathbf{p}} + d_{\mathbf{k}-\mathbf{p}}$. The decay rate displays a complex pattern of peaks and dips. A strong peak near the $\Gamma$ point corresponds to a mirror symmetry of the magnon bands [27]. The other peaks correspond to saddle points of the energy denominator in the self-energy (5). The decay rate of the down band, shown in Fig. 2(b), is determined by the scattering channel $d_{\mathbf{k}} \to d_{\mathbf{p}} + d_{\mathbf{k}-\mathbf{p}}$, and in the vicinity of the K point it is sharply suppressed, i.e. the magnon at the Dirac nodal point is a well-defined excitation whose lifetime shrinks rapidly for small excursions away from the K point.

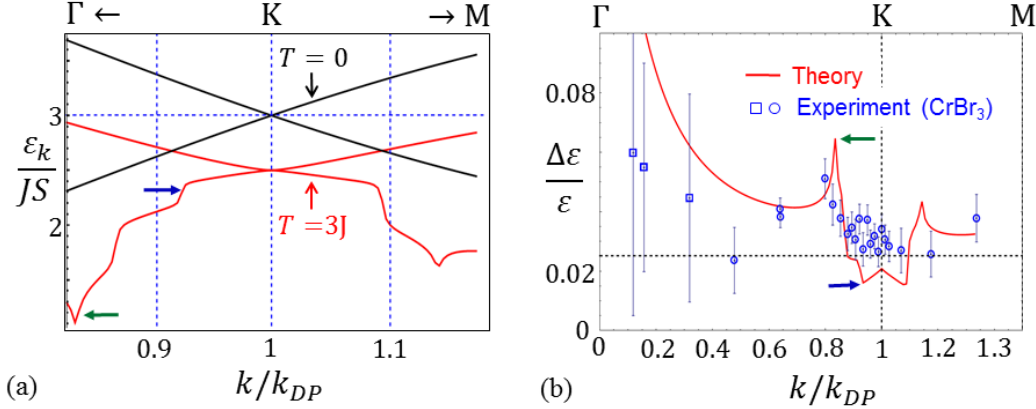

FIG 3. (a) Temperature-dependent renormalization of the Dirac dispersion near K point. The bare and renormalized spectra are shown in red and black lines. (b) Temperature-dependent renormalization of magnon energy spectrum copied from Ref. [2] overlaid with our theory (red). The green and blue arrows in both panels are used as a guide to the eye and point to the corresponding features in the spectrum. In panel (b), the horizontal dashed line represents the energy renormalization due to the constant Hartree term (3). The momentum $k_{DP} = 4\pi/3\sqrt{3}a$ denotes the position of the Dirac point.

We also plot the real part of the self-energy in the bottom row of the Fig 2: in panel (c) for the up band and in panel (d) for the down band. The real and imaginary parts obey the Kramers-Kronig relation, so a peak in the scattering rate corresponds to a feature in the real part of the self-energy. Results in lower panels in Fig. 2, closely correlate with the upper panels. We pay special attention to the $K$ point where the Dirac nodal point is located. Here the values of the real parts of the self-energy for the up and down bands are equal within the numerical precision. The interaction preserves the degeneracy of the Dirac nodal point. To estimate the effect of renormalization on the magnon energy spectrum, we compare the bare spectrum $\varepsilon(\mathbf{k})$ and temperature-dependent renormalized spectrum

$$\varepsilon_T^{u,d}(\mathbf{k}) = \varepsilon^{u,d}(\mathbf{k}) + \Sigma_T^{(1)}(\mathbf{k}) + \Sigma_T^{(2)}(\mathbf{k}) = (1 - \alpha_1 T^2)\varepsilon^{u,d}(\mathbf{k}) + \Sigma_T^{(2)}(\mathbf{k}), \qquad (6)$$

in the vicinity of the $K$ point in momentum space in Fig. 3(a). Interaction leads to renormalization of the energy of the Dirac nodal point together with a decrease of Dirac velocity in the immediate vicinity of the $K$ point. The latter behavior is in the stark contrast to that for the fermionic Dirac materials, where the Coulomb interaction, leads to a logarithmic increase of the Dirac velocity [13, 14]. There are also sharp kinks, marked by the green and blue arrows, in the self-energies, which correspond to equally sharp features in the magnon decay rate and strongly reshape the original linear Dirac dispersion.

We believe that these calculations explain the old unresolved puzzle of the peculiar magnon-band renormalization in CrBr$_3$ observed in [5, 6]. In Fig. 3(b), we plot the calculated self-energy shift together with the experimental points from Ref. [5] for CrBr$_3$. The plot shows a relative renormalization of the magnon dispersion of the down band, which is $\Delta\varepsilon(\mathbf{k})/\varepsilon(\mathbf{k}) = \left[\varepsilon^d_{T=6K}(\mathbf{k}) - \varepsilon^d_{T=20K}(\mathbf{k})\right]/\varepsilon^d_{T=6K}(\mathbf{k})$ in the notations of the current work. In the literature, distinct values for the coupling constant are cited: $J$ = 11 K [28], $J$ = 16 K [6] and $J$ = 28 K [29]. We evaluate $\Delta\varepsilon/\varepsilon$ from our theory for the coupling constant $J$ = 18 K, which best fits the experimental data. We then plot $\Delta\varepsilon/\varepsilon$ with the red line and overlay it on the experimental data in Fig. 3(b). Surprisingly, the theoretical curve and the experimental data agree reasonably well. In particular, the self-energy correction in both experiment and theory undergo rapid evolution in a small region bounded by the peak, marked by the green arrow, near $K$.

The strong energy dependence of the self-energy is the result of in-plane interactions and Dirac spectrum. CrBr$_3$ is a three-dimensional (3D) material composed of stacks of weakly-coupled two-dimensional (2D) honeycomb layers. To address three dimension aspects one would need to include the interlayer coupling. We expect that the interlayer coupling $J_z$ may lead to a weak broadening of peaks in the self energy shown in Fig. 2, but the magnitude of the effect is small since the interlayer coupling J$_z$ is much smaller than the intralayer coupling J$_z$<<J [6].

The results obtained in this work are universal and apply to other types of bosonic Dirac materials. The profile of Re$\Sigma(\mathbf{k})$ in momentum space in Fig. 2 is determined predominantly by energy denominators and less so by the matrix elements in Eqs. (4) and (5). So, the momentum dependence of these quantities could be viewed as a unique signature of the honeycomb dispersion. Thus, our findings should be highly relevant to other bosonic honeycomb systems, e.g. photonic crystals [8, 9, 30, 31, 32, 33, 34].

    IV.    Results – Surface states in a 3D model.

Dirac nodes have profound consequences for the topological surface states that are analogous to Fermi nodal lines. Because the honeycomb lattice is bipartite, it is a primary candidate to possess topological Shockley edge states (otherwise known as the Su-Schrieffer-Heeger states). Crystal structure of CrBr$_3$ shown in Fig.4(a) shows honeycomb Cr layers stacked in the order that is an analogue of the ABC-stacked graphene layers, which have received a lot of attention because of an unusual spectrum and edge states [20, 21, 22]. We assume the simplest model in which the honeycomb layers are coupled via the vertical bonds with a strength $J_z$ as illustrated in Fig. 4(a). So, the corresponding free magnon Hamiltonian becomes

$$H_0(\mathbf{k}) = JS \begin{pmatrix} 3+\gamma_z & -\gamma_\mathbf{k}-\gamma_z e^{ik_z} \\ -\gamma_\mathbf{k}^* - \gamma_z e^{-ik_z} & 3+\gamma_z \end{pmatrix}. \quad (7)$$

By comparing Eq. (7) with Eq. (1) one observes that the matrix elements acquire extra terms proportional to $\gamma_z = J_z/J$. Let us analyze the fate of the Dirac point given by the equation $\gamma_\mathbf{k} + \gamma_z e^{ik_z} = 0$. For small $\gamma_z$, the solution gives helical contours winding around the corners of the Brillouin zone as shown in Fig. 4(b). This nodal line, i.e. a line where the magnon branches are degenerate, retains the topological properties of the usual Dirac point. In particular, the nodal line has a topological Berry phase

$$\oint d\mathbf{k} \langle \psi_\mathbf{k} | i\partial_\mathbf{k} | \psi_\mathbf{k} \rangle = \pi, \quad (8)$$

which leads to flat band surface states (in the fermionic case) given by the projection of the nodal line onto the 2D momentum space [20, 21, 22] as illustrated in Fig. 4(b).

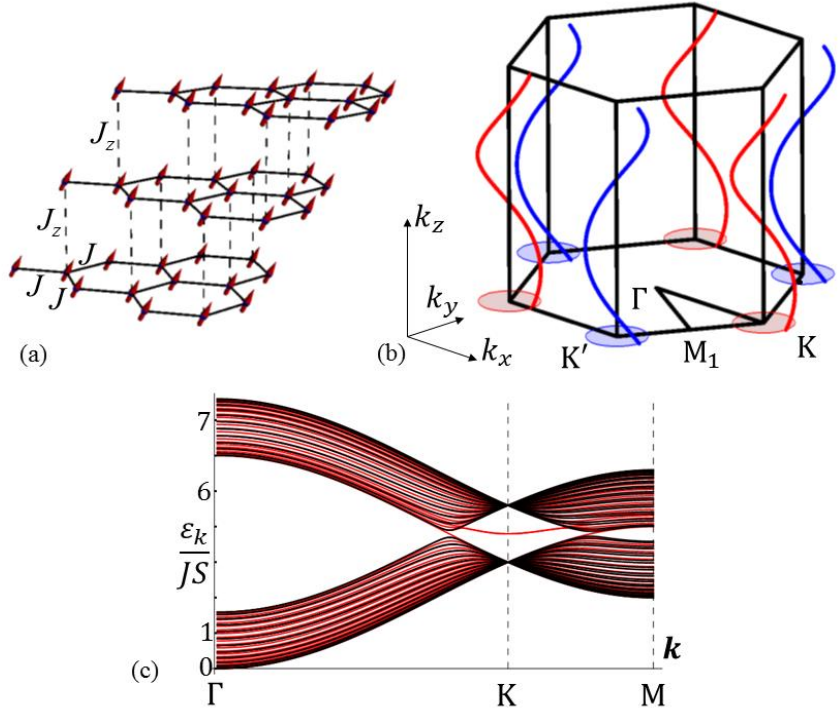

FIG 4. (a) 3D lattice structure of CrBr$_3$. The honeycomb layers are composed of Cr atoms and are stacked in the ABC order. (b) The band structure: Dirac nodal lines wind around the corners of the Brillouin zone. (c) Magnon energy spectrum for a 25 layer slab of a three dimensional material. Results for open vs closed boundary conditions are shown with red and black lines.

Now we evaluate the surface states in the discussed 3D model. Although the surface states are calculated at the single-particle level, it turns out that the surface magnons differ from their fermionic analogues. We compute the magnon dispersion in a slab geometry of N=25 layers and plot the result in Fig. 4(c). The figure contains the spectrum for the closed (in black) and open (in red) boundary conditions. In this representation, the red lines unmatched by the black lines correspond to the surface states.

One can observe the magnon surface states appearing in the vicinity of the K point, which agrees with Fig. 4(b). However, in contrast to what occurs for fermions on stacked honeycomb lattices, the magnon surface states are clearly dispersive. To explain this, we recall the original Heisenberg Hamiltonian $\mathcal{H} = -J \sum_{\langle i,j \rangle} (\mathbf{S}_i \cdot \mathbf{S}_j)$ and the Holstein-Primakoff transformation, e.g. for the z-th component of the spin operators $S_i^z = S - a_i^\dagger a_i$. Observe that *inter-site* coupling in the Heisenberg Hamiltonian generates an *on-site* potential in the bosonic language, i.e. $-JS_i^z S_j^z = -JS^2 + JSa_i^\dagger a_i + JSb_j^\dagger b_j + \mathcal{O}(a^4)$ ($i \neq j$). For the bulk, this procedure generates the diagonal terms in the Hamiltonians (1) and (7). For the surface, inspection of Fig. 4a shows that one sublattice – we assume it is A - will be coupled (to the B sublattice) in the layer immediately below, but not the other. This generates an extra diagonal surface term $H_{surface} = -J_z S a^\dagger a$ for the A but not the B sublattice, in addition to the off-diagonal terms which allow hopping from the A sublattice of the surface to the B sublattice. In the absence of the diagonal surface term, we obtain dispersionless topological surface states that are identical to the fermionic analogues, see e.g. Ref. [22]. Note that the decay length $\xi_k \propto 1/\Delta_\mathbf{k}$ of this surface state $\psi_\mathbf{k}(z) \sim \exp(-z/\xi_\mathbf{k})$ in the z direction is inversely proportional to the bulk gap $\Delta_\mathbf{k}$, where $\mathbf{k} = (k_x, k_y)$ is a good quantum number for a surface cleaved parallel to the layers. Qualitatively, the dispersion of the magnon surface states can then be understood by treating the surface term as a perturbation. Already the first-order perturbation theory produces a **k**-dependent dispersion of the surface states $\varepsilon_{surface}(\mathbf{k}) = \langle \psi_\mathbf{k}(z) | H_{surface} | \psi_\mathbf{k}(z) \rangle$. Indeed, the non-perturbative exact diagonalization of the bosonic Hamiltonian together with the surface term confirms that the surface state acquires a strong dispersion (as shown in the Supplemental Material). We further relax the interplanar Heisenberg coupling $J_z$ as the surface is approached; Fig. 4(c) is plotted for a reduced coupling $J_z^{surface} = 0.5 J_z^{bulk}$ between the surface and penultimate layers. Our discussion extends the previous analysis of the edge states in 2D photonic materials. We also mention that the described mechanism for generating the dispersion of the surface states is universal and should be of relevance to other types of bosonic Dirac materials, e.g. topological phononic and photonic systems [8, 9, 30, 31, 32, 33, 34, 35].

V. Concluding remarks

Noninteracting particles in a 2D honeycomb lattice exhibit Dirac excitations regardless of the statistics. We consider the question of difference between bosonic and fermionic excitations on such lattices. Two key features of the bosonic Dirac material were identified as important: i) Interacting bosons lead to a non-divergent velocity renormalization near the Dirac points. In contrast, Coulomb interactions between fermions lead to a logarithmic correction of the Dirac velocity. ii) non-Bravais nature of the honeycomb lattice structure leads to significant modifications of the spin-wave interactions. More generally, non-Bravais lattices, which entail multicomponent wave-functions for quasiparticles, provide a route to Dirac quasiparticles and distinct surface states. iii) we found topological surface states in $CrBr_3$ that are the bosonic analog of Shockley edge and surface states (also known as the Su-Schrieffer-Heeger states).

To prove the point, we consider the case of ferromagnetic Cr trihalides. Magnetic excitations in honeycomb ferromagnetic layers, the essential element of magnetism in Cr trihalides, possess Dirac nodes. Half a century ago, neutron scattering experiments observed an anomalous momentum-dependent renormalization of the magnon spectrum in $CrBr_3$ near Dirac points [4,5]. By evaluating the self-energy due to the magnon-magnon interactions, we obtained a good agreement with data from Refs. [4,5] and, thus, resolved a fifty-year-old puzzle.

Given the large diversity of transition metal trihalides [29,36], metamaterials and cold atom systems which can implement bosonic Hamiltonians for honeycomb and other non-Bravais lattices, our predictions are highly relevant and can be extensively checked using modern spectroscopic methods. For example, neutron scattering has advanced greatly over the last five decades, and so we anticipate new work to measure the lifetimes as well as self-energies of spin waves for honeycomb magnets. There are also potential applications of the surface states, which we have discovered, to the emerging area of "magnon spintronics" or "magnonics" [37], for which distinct surface spin waves would be advantageous because they avoid bulk dissipation.

Recent experimental observation of the single layer honeycomb ferromagnets [23,24] in $CrI_3$ and other materials highlights the important role the Dirac nodes will play in bosonic Dirac materials realized in Van der Waals magnetic structures.

*Acknowledgment.* This work is supported by US DOE BES E3B700, ERC DM-321031, KAW-2013-0096 and Dr. Max Rössler, the Walter Haefner Foundation and the ETH Zurich Foundation. H. Å. Acknowledges the Knut and Wallenberg foundation for the financial support (Grant No. KAW-2013-0020) and Villum Foundation. We are grateful to A. Chernyshev for useful discussions.

# Supplemental Material

## A. Bosonic Hamiltonian

We follow the standard procedure and bosonize the ferromagnetic Heisenberg Hamiltonian on the honeycomb lattice

$$\mathcal{H} = -J \sum_{\langle i,j \rangle} (\mathbf{S}_i \cdot \mathbf{S}_j). \qquad (9)$$

We introduce the bosonic annihilation operators $a$ and $b$ corresponding to the two sublattices A and B. We relate the spin and boson operators using the Holstein-Primakoff transformation truncated to the first order in $1/S$ as follows

$$\begin{aligned}
S_i^x + iS_i^y &= \sqrt{2S}\left(a_i - \frac{a_i^\dagger a_i a_i}{4S}\right) + \mathcal{O}\left(\frac{1}{S^{3/2}}\right), \\
S_i^x - iS_i^y &= \sqrt{2S}\left(a_i^\dagger - \frac{a_i^\dagger a_i^\dagger a_i}{4S}\right) + \mathcal{O}\left(\frac{1}{S^{3/2}}\right), \qquad (10) \\
S_z &= \hbar\left(S - a_i^\dagger a_i\right),
\end{aligned}$$

(similarly for the B-sublattice). We substitute the leading order terms from the Holstein-Primakoff Eq. (10) in the Hamiltonian (9), do the Fourier transformation $a_j = 1/\sqrt{N}\sum_{\mathbf{k}} e^{i\mathbf{k}\cdot\mathbf{r}_j} a_{\mathbf{k}}$ (similarly for $b_j$) and obtain the linear spin-wave Hamiltonian

$$\mathcal{H} = \sum_{\mathbf{k}} \Psi_{\mathbf{k}}^\dagger H_0(\mathbf{k}) \Psi_{\mathbf{k}}, \qquad H_0(\mathbf{k}) = JS \begin{pmatrix} 3 & -\gamma_{\mathbf{k}} \\ -\gamma_{\mathbf{k}}^* & 3 \end{pmatrix}. \qquad (11)$$

The eigenvalues

$$\varepsilon_{\mathbf{k}}^{u,d} = JS(3 \pm |\gamma_{\mathbf{k}}|) \qquad (12)$$

correspond to the optical and acoustic magnon branches that we also refer to as up and down branches. The corresponding eigenstates are $\Psi_{\mathbf{k}}^{u,d} = \left(e^{i\phi_{\mathbf{k}}/2}, \mp e^{-i\phi_{\mathbf{k}}/2}\right)/\sqrt{2}$. The operators corresponding to the distinct magnon branches $u_{\mathbf{k}}, d_{\mathbf{k}}$ are related to the sublattice annihilation operators $a_{\mathbf{k}}, b_{\mathbf{k}}$ via the unitary transformation

$$\begin{aligned}
u_{\mathbf{k}} &= \frac{1}{\sqrt{2}}\left(e^{-\frac{i\phi_{\mathbf{k}}}{2}} a_{\mathbf{k}} - e^{\frac{i\phi_{\mathbf{k}}}{2}} b_{\mathbf{k}}\right), \\
d_{\mathbf{k}} &= \frac{1}{\sqrt{2}}\left(e^{-\frac{i\phi_{\mathbf{k}}}{2}} a_{\mathbf{k}} + e^{\frac{i\phi_{\mathbf{k}}}{2}} b_{\mathbf{k}}\right).
\end{aligned} \qquad (13)$$

## B. Interactions in a honeycomb ferromagnet

Let us comment on the difference between the honeycomb ferromagnet, considered here, and the honeycomb antiferromagnet. In the case of the ferromagnet, the Hamiltonian commutes with the global spin-rotation operator generator $S^z = \sum_i S_i^z$. This determines the four-magnon interaction in Eq. (2) of the main text.

$$\mathcal{V} = \frac{J}{4N} \sum_{\{\mathbf{k}\}} \gamma^*_{\mathbf{k}_2} a^\dagger_{\mathbf{k}_1} b^\dagger_{\mathbf{k}_2} a_{\mathbf{k}_3} a_{\mathbf{k}_4} + \gamma_{\mathbf{k}_4} a^\dagger_{\mathbf{k}_1} a^\dagger_{\mathbf{k}_2} a_{\mathbf{k}_3} b_{\mathbf{k}_4}$$
$$+ \gamma_{\mathbf{k}_2} b^\dagger_{\mathbf{k}_1} a^\dagger_{\mathbf{k}_2} b_{\mathbf{k}_3} b_{\mathbf{k}_4} + \gamma^*_{\mathbf{k}_4} b^\dagger_{\mathbf{k}_1} b^\dagger_{\mathbf{k}_2} b_{\mathbf{k}_3} a_{\mathbf{k}_4} - 4\gamma_{\mathbf{k}_4-\mathbf{k}_2} a^\dagger_{\mathbf{k}_1} b^\dagger_{\mathbf{k}_2} a_{\mathbf{k}_3} b_{\mathbf{k}_4}. \quad (14)$$

Observe that each interaction term consists of the two creation and annihilation operators, which manifestly conserve the magnon number. The four magnon interaction acts upon the states with at least two excited magnons. Thus, in the limit of zero temperature $T=0$, where no magnons are present, the interaction has no effect. In contrast for finite temperatures, the interaction effects should become more pronounced. However, in the case of the canted antiferromagnet, such a rotational symmetry is absent, which permits the three-magnon terms in the interaction Hamiltonian. The effect of the three-magnon process is non-zero even at absolute zero $T=0$. Thus, although the single-particle spectra may be nominally identical in the cases of the honeycomb ferromagnet and antiferromagnet, the effects of interaction terms are distinct. In what follows, we will describe this finite temperature interaction effects.

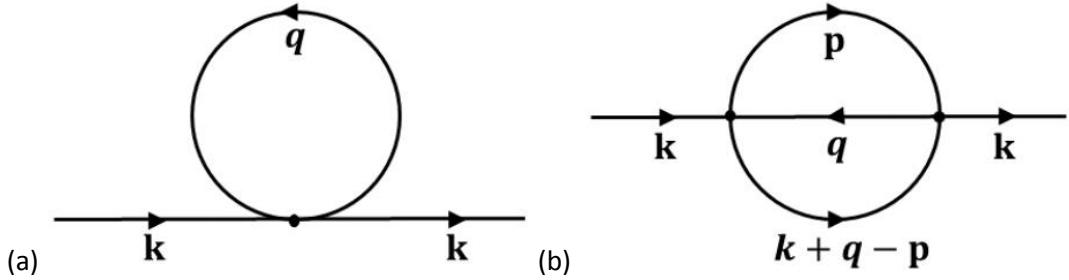

Fig. 1: (a) One- and (b) two-loop contribution to magnon self-energy. Variable *q* denotes the momentum of thermally excited magnons. For small temperatures T<<J, only the low-energy magnons with small *q* are excited.

**B.1. Hartree term**: Here we explain the Hartree energy renormalization in Eq. (3) of the main text according to the diagram shown in Fig. 1(a). The Matsubara summation gives the thermodynamic expectation value of the number of bosons with momentum $q$ according to the Bose-Einstein distribution $f(\varepsilon_q) = 1/[\exp(\varepsilon_q/T) - 1]$. For small temperatures $T \ll J$, only the low-energy down-magnons with momenta close to $q = 0$ are excited. In order to calculate the self-energy, we replace the product of two operators by their expectation values. There are four ways to accomplish this:

$$\gamma^*_{\mathbf{k}_2} a^\dagger_{\mathbf{k}_1} b^\dagger_{\mathbf{k}_2} a_{\mathbf{k}_3} a_{\mathbf{k}_4} \to \gamma^*_{\mathbf{k}_2} \left[ a^\dagger_{\mathbf{k}_1} \langle b^\dagger_{\mathbf{k}_2} a_{\mathbf{k}_3} \rangle a_{\mathbf{k}_4} + a^\dagger_{\mathbf{k}_1} \langle b^\dagger_{\mathbf{k}_2} a_{\mathbf{k}_4} \rangle a_{\mathbf{k}_3} + b^\dagger_{\mathbf{k}_2} \langle a^\dagger_{\mathbf{k}_1} a_{\mathbf{k}_3} \rangle a_{\mathbf{k}_4} + b^\dagger_{\mathbf{k}_2} \langle a^\dagger_{\mathbf{k}_1} a_{\mathbf{k}_4} \rangle a_{\mathbf{k}_3} \right] \quad (15)$$

So, we contract the pairs of operators in Eq. (14) of the main text using the expectation value $\langle d^\dagger_\mathbf{q} d_\mathbf{q} \rangle = f(\varepsilon^d_\mathbf{q})$ and the wave functions defined in section I. This produces the following expectation values in Eq. (15),

$$\begin{aligned}
\langle a^\dagger_\mathbf{q} a_\mathbf{q} \rangle &= \langle b^\dagger_\mathbf{q} b_\mathbf{q} \rangle = \frac{1}{2} f(\varepsilon^d_\mathbf{q}), \\
\langle a^\dagger_\mathbf{q} b_\mathbf{q} \rangle &= \langle b^\dagger_\mathbf{q} a_\mathbf{q} \rangle^* = \frac{e^{-i\varphi_\mathbf{q}}}{2} f(\varepsilon^d_\mathbf{q}),
\end{aligned} \quad (16)$$

So, the Hartree contribution can be rewritten as,

$$H_{\text{Hartree}} = \sum_\mathbf{k} h(T) a^\dagger_\mathbf{k} a_\mathbf{k} + h(T) b^\dagger_\mathbf{k} b_\mathbf{k} + g_\mathbf{k}(T) a^\dagger_\mathbf{k} b_\mathbf{k} + g^*_\mathbf{k}(T) b^\dagger_\mathbf{k} a_\mathbf{k}, \quad (17)$$

where for brevity we define the coefficients,

$$\begin{aligned}
h(T) &= \frac{J}{2N} \sum_\mathbf{q} (|\gamma_\mathbf{q}| - |\gamma_0|) f(\varepsilon^d_\mathbf{q}), \\
g_\mathbf{k}(T) &= \frac{J}{2N} \sum_\mathbf{q} (\gamma_\mathbf{k} - e^{i\phi_q} \gamma_{\mathbf{k}-\mathbf{q}}) f(\varepsilon^d_\mathbf{q}).
\end{aligned} \quad (18)$$

For small temperature $T \ll J$, it is valid to expand Eq. (18) in powers of $\mathbf{q}$ and integrate. So, we obtain, $h(T) = -T^2 \frac{\pi}{8\sqrt{3}JS^2}$ and $g_\mathbf{k}(T) = -T^2 \frac{\pi}{24\sqrt{3}JS^2} \Delta\gamma_\mathbf{k}$. Given that $\Delta\gamma_\mathbf{k} = -\gamma_\mathbf{k}$, we find that $g_\mathbf{k}(T) = -\frac{h(T)\gamma_\mathbf{k}}{3}$. So, the remaining uncontracted operators give the following self-energy,

$$\Sigma_1(\mathbf{k}) = -\alpha T^2 JS \begin{pmatrix} 3 & -\gamma_\mathbf{k} \\ -\gamma^*_\mathbf{k} & 3 \end{pmatrix} = -\alpha T^2 H_0(\mathbf{k}), \quad \alpha = \frac{\pi}{24\sqrt{3}J^2 S^3}, \quad (19)$$

written in the sublattice basis. We compare the free Hamiltonian (11) with the Hartree self-energy (19) and observe the same matrix structure. So, the renormalized Hamiltonian $H_0(\mathbf{k}) + \Sigma_1(\mathbf{k})$ has the same wave functions as the linear spin wave Hamiltonian and the renormalized energy spectrum is $(1-\alpha T^2)\varepsilon_{\mathbf{k}}^{u,d}$, where $\varepsilon_{\mathbf{k}}^{u,d}$ is the energy spectrum of the free Hamiltonian (11). We observe that the Hartree term leads to a temperature-dependent rescaling of both the branches of the energy spectrum uniformly throughout the Brillouin zone.

**B.2. Rainbow diagram:** Next, we analyze the effect of interaction to the next order in $\frac{1}{S}$. The self-energy corresponding to the diagram in Fig. 1(b) is

$$\Sigma_2(i\omega,\mathbf{k}) = \frac{1}{N^2\beta^2}\sum_{\mathbf{q},\mathbf{p},\omega_i} \frac{|V_{\mathbf{k},\mathbf{q};\mathbf{p}}|^2}{(i\omega_1 - \varepsilon_{\mathbf{p}})(i\omega_2 - \varepsilon_{\mathbf{k}+\mathbf{q}-\mathbf{p}})(i(\omega_1+\omega_2-\omega) - \varepsilon_{\mathbf{q}})}. \quad (20)$$

We perform the Matsubara summation and obtain

$$\Sigma_2(\omega,\mathbf{k}) = \frac{1}{N^2}\sum_{\mathbf{q},\mathbf{p}} \frac{|V_{\mathbf{k},\mathbf{q},\mathbf{p}}|^2 F_{\mathbf{k},\mathbf{q},\mathbf{p}}}{\omega + \varepsilon_{\mathbf{q}} - \varepsilon_{\mathbf{p}} - \varepsilon_{\mathbf{k}+\mathbf{q}-\mathbf{p}} + i\delta},$$
$$F_{\mathbf{k},\mathbf{q};\mathbf{p}} = \left[1+f(\varepsilon_{\mathbf{p}})\right]\left[1+f(\varepsilon_{\mathbf{k}+\mathbf{q}-\mathbf{p}})\right]f(\varepsilon_{\mathbf{q}}) - f(\varepsilon_{\mathbf{p}})f(\varepsilon_{\mathbf{k}+\mathbf{q}-\mathbf{p}})\left[1+f(\varepsilon_{\mathbf{q}})\right]. \quad (21)$$

Note that the self-energy is a complex-valued function, where both the real and imaginary parts have a physical significance. The imaginary part of the diagram gives the scattering rate of the magnon excitations $W_{\mathbf{k}} = -\text{Im}\Sigma(\varepsilon_{\mathbf{k}},\mathbf{k})$, whereas the real part determines the renormalization of the magnon energy spectrum $\Delta\varepsilon_{\mathbf{k}} = \text{Re}\Sigma(\varepsilon_{\mathbf{k}},\mathbf{k})$.

Equation (21) can be calculated using a numerically heavy integration in the 4D momentum space $(\mathbf{p},\mathbf{q}) = (p_x, p_y, q_x, q_y)$. However, we choose a simplifying route and notice that Eq. (21) can be further reduced in the limit of small temperatures $T \ll J$. We observe that the first term in the thermodynamic function $F_{\mathbf{k},\mathbf{q};\mathbf{p}}$ corresponds to the direct scattering process, i.e. where the original magnon with momentum $\mathbf{k}$ and the thermal magnon with momentum $\mathbf{q}$ scatter into magnons with momenta $\mathbf{p}$ and $\mathbf{k}+\mathbf{q}-\mathbf{p}$, whereas the second term to the reverse scattering process. For small temperatures $T \ll J$, the direct scattering process dominates, and $F_{\mathbf{k},\mathbf{q};\mathbf{p}}$ can be approximated as $F_{\mathbf{k},\mathbf{q};\mathbf{p}} \approx f(\varepsilon_{\mathbf{q}}^d)$, which represents the occupation number of the thermal "down" band magnons of momentum $\mathbf{q}$. As temperature is lowered, only the magnons from the "down" band with momenta close to $\Gamma$ point are excited, i.e. $\mathbf{q}$ is small. We further notice that the matrix element vanishes for $V_{\mathbf{k},\mathbf{q};\mathbf{p}}^\alpha|_{\mathbf{q}=0} = 0$ (at least for the on-shell processes). Thus the expansion of the matrix element starts with the linear order in $\mathbf{q}$: $V_{\mathbf{k},\mathbf{q};\mathbf{p}} = \mathbf{q}\cdot\mathbf{v}_{\mathbf{k};\mathbf{p}} + \mathcal{O}(q^2)$. This expansion together with the simplification $F_{\mathbf{k},\mathbf{q};\mathbf{p}} \approx f(\varepsilon_{\mathbf{q}}^d)$ allows us to integrate over $\mathbf{q}$ and obtain

$$\Sigma_2(\omega,\mathbf{k}) = \alpha T^2 A \int \frac{d^2 p}{(2\pi)^2} \frac{|\mathbf{v}_{\mathbf{k};\mathbf{p}}|^2}{\omega - \varepsilon_{\mathbf{p}} - \varepsilon_{\mathbf{k}-\mathbf{p}} + i\delta}, \quad (22)$$

where $A = \frac{3\sqrt{3}}{2}$ is the area of the honeycomb unit cell (in the units where interatomic distance a=1). To sum up, we integrated over the thermal magnons $\mathbf{q}$ analytically, which produced an overall temperature dependence. The power of the temperature dependence is determined by the dimensionality. The remaining integration over the 2D momentum space $\mathbf{p} = (p_x, p_y)$ can now be evaluated numerically. Finally, let us comment on another intricate issue of evaluating the matrix element expansion $\mathbf{v}_{\mathbf{k};\mathbf{p}}$. This expansion is obtained on an isoenergetic surface $0 = E_{\mathbf{k},\mathbf{q};\mathbf{p}} = \Omega + \varepsilon_{\mathbf{q}} - \varepsilon_{\mathbf{p}} - \varepsilon_{\mathbf{k}+\mathbf{q}-\mathbf{p}}$, so the correct equation for the matrix element expansion is the following

$$\mathbf{v}_{\mathbf{k},\mathbf{p}} = \nabla_q V - \nabla_q E \frac{(\nabla_p V \cdot \nabla_p E)}{(\nabla_p E)^2}. \quad (23)$$

### B.3. Matrix elements:

Next, we give the matrix elements obtained for the model (14). The interaction term in Eq. (14) can be written in terms of $u_k$ and $d_k$ using Eq. (13)

$$\mathcal{V} = \frac{1}{N} \sum_{\{\mathbf{k},\mathbf{q},\mathbf{p}\}} \left[ V^{(1)}_{\mathbf{k},\mathbf{q},\mathbf{p}} u^\dagger_\mathbf{p} d^\dagger_{\mathbf{k}+\mathbf{q}-\mathbf{p}} u_\mathbf{k} d_\mathbf{q} + V^{(2)}_{\mathbf{k},\mathbf{q},\mathbf{p}} d^\dagger_\mathbf{p} d^\dagger_{\mathbf{k}+\mathbf{q}-\mathbf{p}} u_\mathbf{k} d_\mathbf{q} + V^{(3)}_{\mathbf{k},\mathbf{q},\mathbf{p}} d^\dagger_\mathbf{p} d^\dagger_{\mathbf{k}+\mathbf{q}-\mathbf{p}} d_\mathbf{k} d_\mathbf{q} + \text{rest} \right] \quad (24)$$

Now, we list the forms of $V^{(1)}_{\mathbf{k},\mathbf{q},\mathbf{p}}$, $V^{(2)}_{\mathbf{k},\mathbf{q},\mathbf{p}}$ and $V^{(3)}_{\mathbf{k},\mathbf{q},\mathbf{p}}$ for three different scattering processes $u_\mathbf{k} + d_\mathbf{q} \to u_\mathbf{p} + d_{\mathbf{k}+\mathbf{q}-\mathbf{p}}$, $u_\mathbf{k} + d_\mathbf{q} \to d_\mathbf{p} + d_{\mathbf{k}+\mathbf{q}-\mathbf{p}}$ and $d_\mathbf{k} + d_\mathbf{q} \to d_\mathbf{p} + d_{\mathbf{k}+\mathbf{q}-\mathbf{p}}$ in Eq. (24). In the scattering processes above, there is always a thermal "down" band magnon, which we label by the incoming momentum by $\mathbf{q}$. The corresponding matrix elements are

$$V^{(1)}_{\mathbf{k},\mathbf{q},\mathbf{p}} = \frac{J}{4} \left( |\gamma_\mathbf{q}| - |\gamma_\mathbf{k}| - |\gamma_\mathbf{p}| + |\gamma_{\mathbf{k}+\mathbf{q}-\mathbf{p}}| \right) \cos \Phi + \frac{J}{2} |\gamma_{\mathbf{q}-\mathbf{p}}| \cos \Xi - \frac{J}{2} |\gamma_{\mathbf{p}-\mathbf{k}}| \cos \Psi,$$

$$V^{(2)}_{\mathbf{k},\mathbf{q},\mathbf{p}} = \frac{iJ}{4} \left( |\gamma_\mathbf{k}| - |\gamma_\mathbf{q}| - |\gamma_\mathbf{p}| - |\gamma_{\mathbf{k}+\mathbf{q}-\mathbf{p}}| \right) \sin \Phi + \frac{iJ}{2} |\gamma_{\mathbf{p}-\mathbf{q}}| \sin \Xi + \frac{iJ}{2} |\gamma_{\mathbf{p}-\mathbf{k}}| \sin \Psi, \quad (25)$$

$$V^{(3)}_{\mathbf{k},\mathbf{q},\mathbf{p}} = \frac{J}{4} \left( |\gamma_\mathbf{k}| + |\gamma_\mathbf{q}| + |\gamma_\mathbf{p}| + |\gamma_{\mathbf{k}+\mathbf{q}-\mathbf{p}}| \right) \cos \Phi - \frac{J}{2} |\gamma_{\mathbf{p}-\mathbf{q}}| \cos \Xi - \frac{J}{2} |\gamma_{\mathbf{p}-\mathbf{k}}| \cos \Psi,$$

where the different angles that enter into the Eq. (25) are $\Phi(\mathbf{k},\mathbf{q},\mathbf{p}) = \frac{\phi_\mathbf{p} + \phi_{\mathbf{k}+\mathbf{q}-\mathbf{p}} - \phi_\mathbf{q} - \phi_\mathbf{k}}{2}$, $\Xi(\mathbf{k},\mathbf{q},\mathbf{p}) = \frac{-\phi_\mathbf{p} + \phi_{\mathbf{k}+\mathbf{q}-\mathbf{p}} + \phi_\mathbf{q} - \phi_\mathbf{k} + 2\phi_{\mathbf{p}-\mathbf{q}}}{2}$ and $\Psi(\mathbf{k},\mathbf{q},\mathbf{p}) = \frac{\phi_\mathbf{p} - \phi_{\mathbf{k}+\mathbf{q}-\mathbf{p}} + \phi_\mathbf{q} - \phi_\mathbf{k} + 2\phi_{\mathbf{k}-\mathbf{p}}}{2}$. The expansion of the matrix elements according to Eq. (23) gives

$$v^{(1)}_{\mathbf{k};\mathbf{p}} = \frac{J}{2} \operatorname{Re} \left[ e^{i\Phi(\mathbf{k},\mathbf{p})} \left( e^{-i\phi_{\mathbf{k}-\mathbf{p}}} \nabla \gamma_{\mathbf{k}-\mathbf{p}} - e^{-i\phi_\mathbf{p}} \nabla \gamma_\mathbf{p} \right) \right],$$

$$v^{(2)}_{\mathbf{k};\mathbf{p}} = -\frac{iJ}{2} \operatorname{Im} \left[ e^{i\Phi(\mathbf{k},\mathbf{p})} \left( e^{-i\phi_{\mathbf{k}-\mathbf{p}}} \nabla \gamma_{\mathbf{k}-\mathbf{p}} + e^{-i\phi_\mathbf{p}} \nabla \gamma_\mathbf{p} \right) \right], \quad (26)$$

$$v^{(3)}_{\mathbf{k};\mathbf{p}} = \frac{J}{2} \operatorname{Re} \left[ e^{i\Phi(\mathbf{k},\mathbf{p})} \left( e^{-i\phi_{\mathbf{k}-\mathbf{p}}} \nabla \gamma_{\mathbf{k}-\mathbf{p}} + e^{-i\phi_\mathbf{p}} \nabla \gamma_\mathbf{p} \right) \right],$$

where $\Phi(\mathbf{k};\mathbf{p}) = \frac{\phi_\mathbf{p} + \phi_{\mathbf{k}-\mathbf{p}} - \phi_\mathbf{k}}{2}$.

### B.4. Scattering rate:

First, let us discuss the details of evaluating the scattering rate. The scattering rate is related to the imaginary part of the self-energy, and we thus obtain from Eq. (22)

$$W_{\mathbf{k}} = \alpha \pi T^2 \int d^2 p \, |\mathbf{v}_{\mathbf{k};\mathbf{p}}|^2 \delta(\varepsilon_{\mathbf{k}} - \varepsilon_{\mathbf{p}} - \varepsilon_{\mathbf{k}-\mathbf{p}}), \qquad (27)$$

The integrand in Eq. (27) contains two terms: the matrix element and the delta function. The delta function determines the kinematically allowed phase space. Since $\mathbf{k}$ is an external parameter, the argument of the delta function generically determines a one-dimensional (1D) contour in the two-dimensional (2D) momentum $\mathbf{p}$ space. The competition between the matrix element and the kinematic factor, given by the delta-function, determines the overall behavior of the scattering rate (27). To give an idea about the kinematic factor, we also compute the scattering density of states

$$K_{\mathbf{k}} = \int d^2 p \, \delta(\varepsilon_{\mathbf{k}} - \varepsilon_{\mathbf{p}} - \varepsilon_{\mathbf{k}-\mathbf{p}}), \qquad (28)$$

which retains only the kinematic part of Eq. (27). Then, one can compare the relative contribution of the matrix element and the kinematic terms by superposing Eqs. (27) and (28).

The bare original energy spectrum contains the "up" and "down" branches, which we label as $u$ and $d$ for brevity. There are multiple scattering channels allowed by energy conservation that we represent as: $u \to u + d$, $u \to d + d$ and $d \to d + d$ (here we omit a label for the "ghost" thermal magnons from the "down" branch). We evaluate numerically the scattering rate Eq. (27) and density of states Eq. (28) by approximating the delta-function with the Lorentzian of finite width $J/100$. The results for the three different processes are shown in Fig. 2(a,c,e) as well as within the main text, along the line $\Gamma \to K \to M \to \Gamma$ in the momentum space. For clarification we also plot the phase space contours for all these three processes in Fig. 2 (b), (d) and (f). Below, we briefly describe each of the three scattering processes in details.

**B.4.1. Process $\mathbf{u} \to \mathbf{u} + \mathbf{d}$**: We give the corresponding plots in panels (a) and (b) if Fig. 2. The scattering rate, shown in panel (a), displays a series of van-Hove-like peaks and dips as a function of momentum $k$. In order to analyze this pattern, we consider the energy conservation corresponding to the process $u \to u + d$

$$0 = \varepsilon_{\mathbf{k}}^u - \varepsilon_{\mathbf{p}}^u - \varepsilon_{\mathbf{k}-\mathbf{p}}^d = JS(|\gamma_{\mathbf{k}}| - 3 - |\gamma_{\mathbf{p}}| + |\gamma_{\mathbf{k}-\mathbf{p}}|), \qquad (29)$$

which is also the argument of delta function in Eqs. (27)-(28). Recall that $k$ denotes the momentum of the original magnon, whereas $p$ - the momentum of the one of the two new magnons, and the momentum of another $k$-$p$ is determined by the momentum conservation. Solution of equation (29) with respect to the momentum $p$ determines scattering contours in the 2D momentum $p$ space, which we show in the panel (b). For reference, the coloring scheme in panels (a) and (b) is identical: the vertical line of a particular color in panel (a) corresponds to the momentum $k$, for which the scattering contour (29) of the same color is plotted in panel (b). For instance, consider the peak in the scattering rate located halfway between the $\Gamma$ and K points in the panel (a). In panels (b), we plot the scattering contours in red and green corresponding to the momenta $k$, in the vicinity the scattering rate peak as shown by the red and green vertical line in panel (a). So, by comparing panels (a) and (b), we observe that the van-Hove peak in the scattering rate corresponds to the reconnection of the scattering contours (compare red and green lines). With further increase of the momentum $k$ towards the K point, the scattering contour shrinks around the K point as shown in blue and magenta lines in panel (b). Therefore, the scattering density of states decreases and ultimately vanishes for $k$ = K. The scattering rate has the strongest singularity for k near the $\Gamma$ point. Notice that exactly at $\Gamma$, i.e. for k = 0, Eq. (18) is trivially satisfied for all $p$ within the Brillouin zone. The apparent divergence of the scattering rate stems from the mirror symmetry of the magnon spectrum [27][5].

**B.4.2. Process** $\mathbf{u} \to \mathbf{d} + \mathbf{d}$: The corresponding plots are shown in Fig. 2(c) and (d). The argument of the delta function in Eqs. (27)-(28), gives the energy conservation for this scattering channel

$$0 = \varepsilon_{\mathbf{k}}^{u} - \varepsilon_{\mathbf{p}}^{d} - \varepsilon_{\mathbf{k-p}}^{d} = JS\left(|\gamma_{\mathbf{k}}| - 3 + |\gamma_{\mathbf{p}}| + |\gamma_{\mathbf{k-p}}|\right), \qquad (30)$$

Because of the energy conservation (30), the process is allowed only for k sufficiently close to the corners K of the Brillouin zone. In particular, for momentum $\mathbf{k}$ close to the $\Gamma$ point, there is no solution of Eq. (30). The scattering rate in panel (b) consists of three characteristic peaks. The side peaks correspond to the van-Hove singularity where the single-particle energy $\varepsilon_{\mathbf{k}}^{u}$ enters the two-magnon continuum $\varepsilon_{\mathbf{p}}^{d} + \varepsilon_{\mathbf{k-p}}^{d}$.

Both processes $u \to u + d$ and $u \to d + d$ determine the scattering rate of the magnons in the up-branch. Figure 2(a) in the main text gives a total decay rate and is a sum over the decay channels discussed above.

**B.4.3. Process** $\mathbf{d} \to \mathbf{d} + \mathbf{d}$: This process happens entirely within the down-band. The corresponding plots are shown in Fig. 2(e, f). The argument of the delta function in Eq. (27) and Eq. (28) gives the energy conservation for this scattering as follows,

$$0 = \varepsilon_{\mathbf{k}}^{d} - \varepsilon_{\mathbf{p}}^{d} - \varepsilon_{\mathbf{k-p}}^{d} = JS\left(-|\gamma_{\mathbf{k}}| - 3 + |\gamma_{\mathbf{p}}| + |\gamma_{\mathbf{k-p}}|\right), (31)$$

The pattern in panel 2(e) has a number of peaks in the vicinity of the corners of the Brillouin zone K. All peaks correspond to reconnections of the scattering contours as shown in panel (e). In the vicinity of the $\Gamma$ point, the dimensionless scattering rate $W_{\mathbf{k}}$ is suppressed, although the corresponding scattering density of states $K_{\mathbf{k}}$ is finite. Physically this means that interactions keep the Goldstone "down" magnons well defined sharp excitations with a long lifetime.

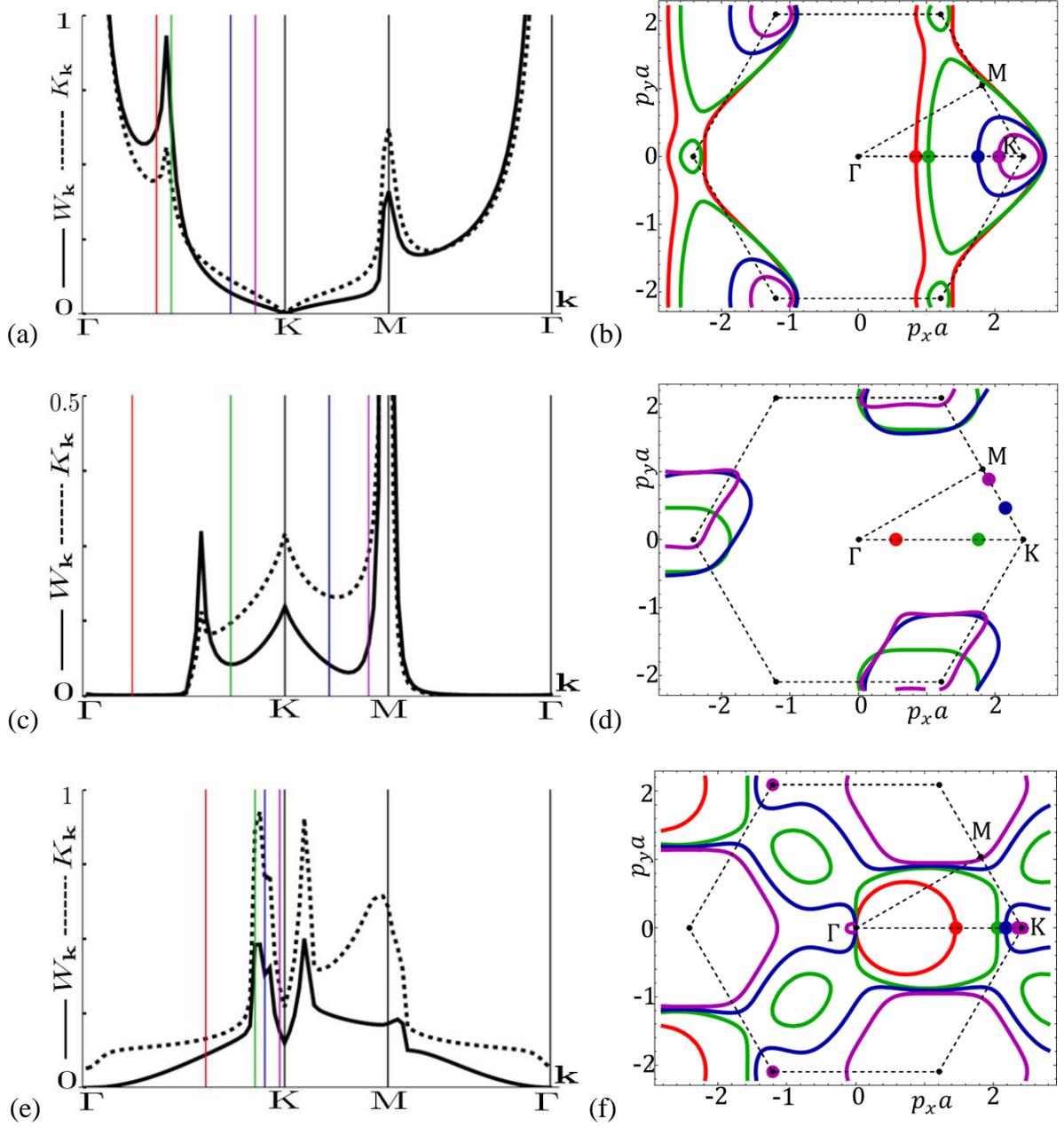

Fig. 2: Left panels: Scattering rate (27) shown with solid and scattering density of states (28) shown with dashed line for **k** along the $\Gamma \to K \to M \to \Gamma$ line in the Brillouin zone. Right column: Scattering contours, i.e. the solutions of Eqs. (29)-(31) with respect to momentum **p**. The top, middle, and bottom rows correspond to different scattering channels $u \to u + d$, $u \to d + d$, and $d \to d + d$ correspondingly. Coloring scheme: Scattering contours of different colors in the right panels correspond to the initial magnon **k** shown with vertical lines of the same color in the left panels.

## B.5. Real part of the self-energy:

The real part of the self-energy also bares a physical significance and gives an interaction-induced renormalization of the magnon energies. So, we calculate the real part of the self-energy (22)

$$\text{Re}\Sigma_2(\varepsilon_\mathbf{k},\mathbf{k}) = \alpha T^2 \sum_\mathbf{p} |\mathbf{v}_{\mathbf{k};\mathbf{p}}|^2 \frac{(\varepsilon_\mathbf{k} - \varepsilon_\mathbf{p} - \varepsilon_{\mathbf{k}-\mathbf{p}})}{(\varepsilon_\mathbf{k} - \varepsilon_\mathbf{p} - \varepsilon_{\mathbf{k}-\mathbf{p}})^2 + \delta^2}, \qquad (32)$$

where $\delta = J/100$ is chosen finite for numerical reasons. In contrast with the imaginary part given only by the on-shell processes, we now consider both the on-shell and off-shell processes. So, six process are considered in calculating the real part of the self-energy. The three processes $u \to u+u$, $u \to u+d$ and $u \to d+d$ contribute to the renormalization of the up-band and the three processes $d \to u+u$, $d \to u+d$ and $d \to d+d$ contribute to the renormalization of the down-band.

Using Eq. (24) and the derived expressions for the matrix elements $\mathbf{v}_{\mathbf{k},\mathbf{p}}$, we evaluate the real part of the self-energy. We consistently find that the self-energy of the down band is negative for momentum $\mathbf{k}$ close to the $\Gamma$ point. This observation makes sense from the point of view of the text-book second-order perturbation theory, which always produces a negative shift of the lowest energy state. This also means that the renormalized energy spectrum of the Goldstone modes is negative, which is clearly unphysical. The only way to fix this inconsistency would be to force the momentum-dependent matrix element $\mathbf{v}_{\mathbf{k},\mathbf{p}}$ to zero where the momenta are close to $\Gamma$ point. In fact, Dyson [3, 4] argued that the Holstein-Primakoff approach incorrectly estimates the matrix elements in the long-wavelength limit. In order to bypass this difficulty, we are motivated by the decay rate shown in Fig. 2(left panels), where we observe that the results with (solid curve) and without (dashed curve) matrix element are similar. So, the role of the matrix element is relatively weak: it is just a well-behaved function that only slightly modifies the results. The gross features of the tunneling spectrum such as the van-Hove peaks are predominantly determined by the energy denominator. So, with these reservations in mind, we suggest that the real part would also be weakly dependent on the matrix element. Given that we cannot extract the correct long-wavelength behavior of the matrix elements from the Holstein-Primakoff transformations anyway, we postulate the following ansatz for all matrix elements

$$|\mathbf{v}_{\mathbf{k},\mathbf{p}}| = [1-\exp(|\mathbf{k}|/\lambda)][1-\exp(|\mathbf{p}|/\lambda)][1-\exp(|\mathbf{k}-\mathbf{p}|/\lambda)], \qquad (33)$$

where $\lambda$ is a constant. Next, we substitute Eq. (33) in Eq. (24) and evaluate the real part of self-energy and obtain the results in Figs. 2 and 3 of the main text.

**C. Comparison with a triangular ferromagnet:**

Because of the two-sublattice (non-Bravais) structure of the honeycomb lattice, there are multiple scattering channels that we have discussed so far. It is also helpful to mention an analogous scattering rate in the triangular lattice (note that the honeycomb lattice is composed of the two interpenetrating triangular lattices). The discussion closely follows that of the honeycomb lattice: we start with the Heisenberg model, bosonize it, and evaluate the free Hamiltonian

$$\mathcal{H}_0 = JS \sum_{\langle ij \rangle} \left( a_i a_j^\dagger - a_i^\dagger a_i + h.c. \right) = \sum_{\mathbf{k}} \varepsilon_{\mathbf{k}} a_{\mathbf{k}}^\dagger a_{\mathbf{k}} \quad (34)$$

as well as the interaction

$$\mathcal{H}_{int} = J \sum_{\langle ij \rangle} \left( a_j^\dagger a_j^\dagger a_j a_i + a_i^\dagger a_j^\dagger a_i a_i + a_i^\dagger a_j^\dagger a_j a_j + a_i^\dagger a_i^\dagger a_i a_j - a_i^\dagger a_j^\dagger a_i a_j \right) \quad (35)$$

In contrast with the honeycomb lattice, there is now only a single magnon branch with spectrum $\varepsilon_{\mathbf{k}} = JS(6 - 2f_{\mathbf{k}})$, where $f_{\mathbf{k}} = \left( \cos(a k_x) + \cos(\frac{a}{2}(k_x + \sqrt{3} k_y)) + \cos(\frac{a}{2}(k_x - \sqrt{3} k_y)) \right)$ and $a$ is the lattice constant. The spectrum has only one branch and is shown in Fig. 3(a). As above, we perform the low-temperature approximation and evaluate the scattering rate (27) as well as the density of states (28) in Fig. 3(b) (to be compared with Fig 2(e)). The decay spectra in the vicinity of the $\Gamma$ point are similar for the triangular and honeycomb lattices as expected: the long-wavelength magnons cannot resolve an internal lattice structure of the lattice. On the other hand, the spectra are quite different close to the Brillouin zone boundary near the K and M points. One similar feature is that both the spectra have a number of peaks near the high-symmetry points. The evolution of the scattering contours of the triangular lattice is shown in Fig. 3(d). We plot the real part of the self-energy in Fig. 3(c).

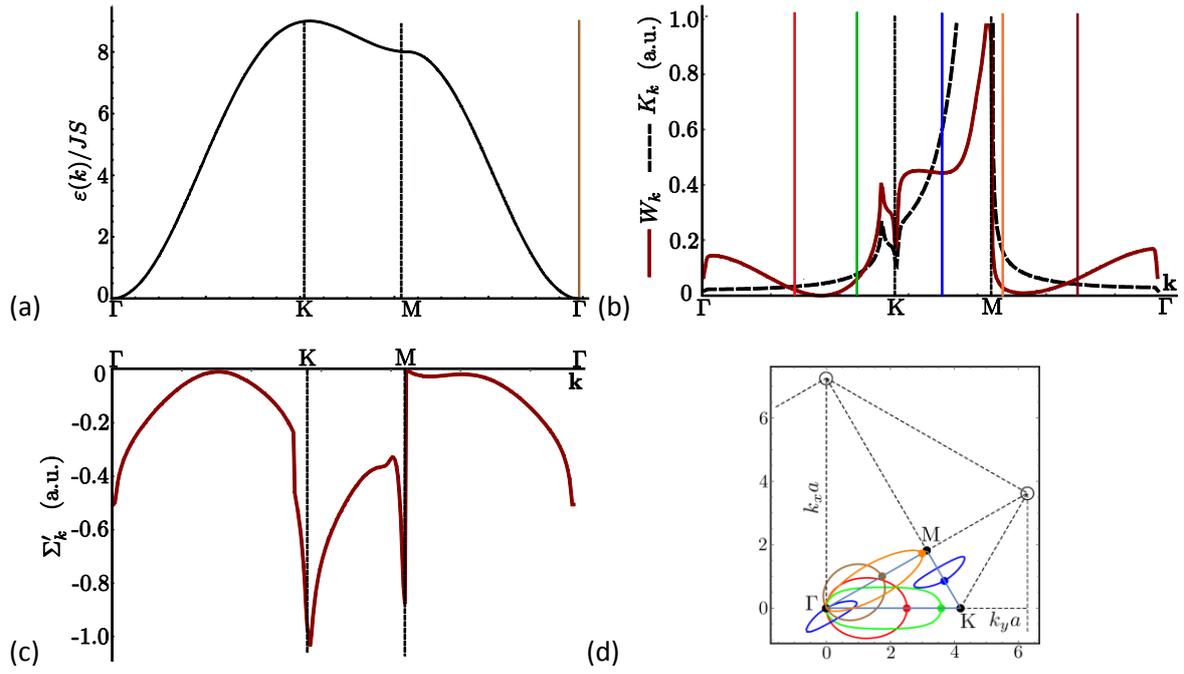

Fig. 3: (a) Magnon dispersion in a triangular ferromagnet. (b) Scattering rate (27) and scattering density of states (28) shown with solid and dashed lines correspondingly. (c) The real part of the self-energy. (c) For each initial momentum **k** shown with a thick dot, we plot a scattering contour with a line of the corresponding color.

## D. Edge states and surface states.

### D.1. Edge states in 2D.

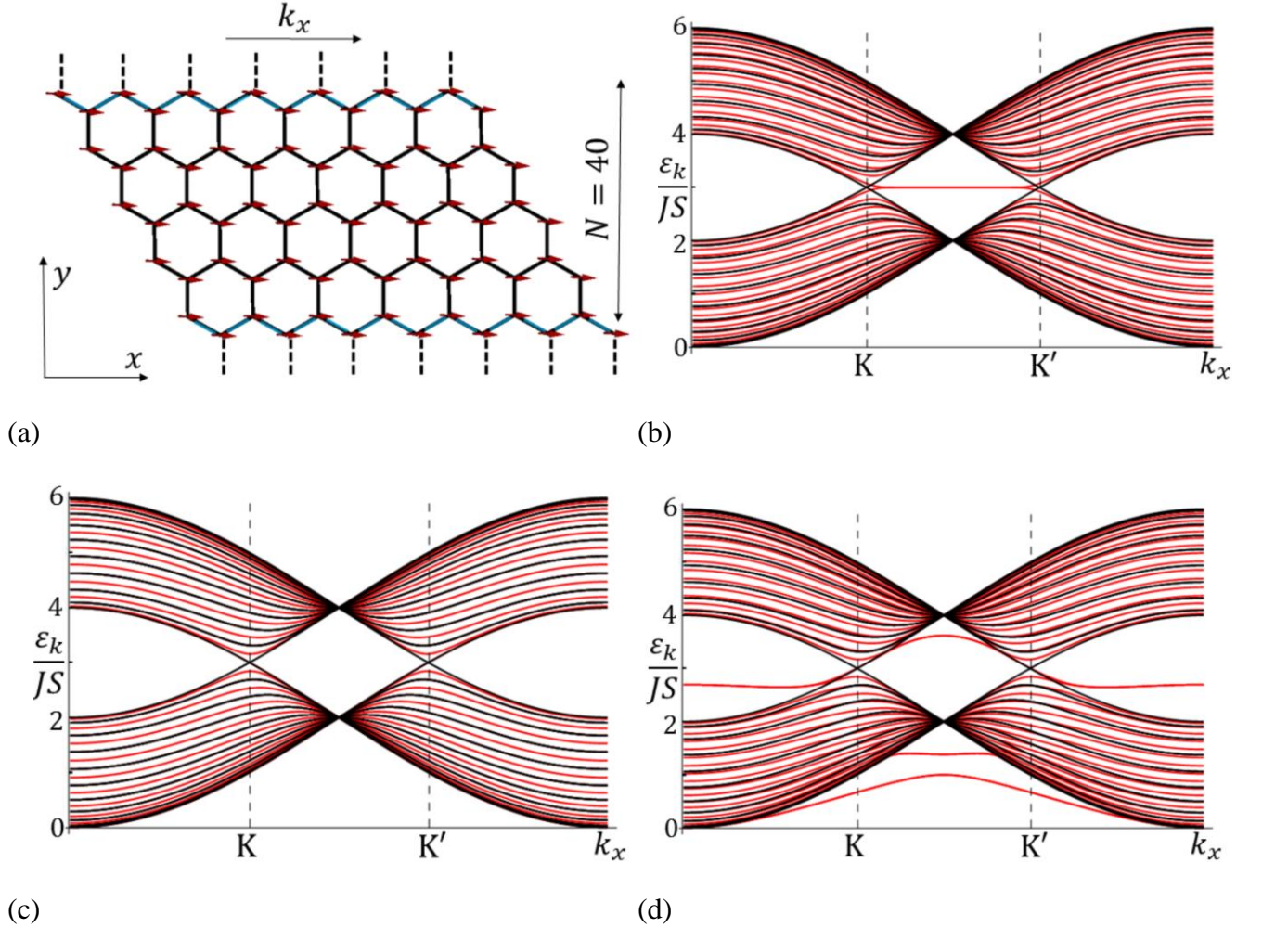

(a)  (b)  (c)  (d)

Fig 4: (a) Geometry: Ribbon of a finite width N = 40 atoms and infinite in x direction. (b-d) Magnon energy spectrum (red) for various boundary conditions. (b) Spectrum (red) for the truncated Hamiltonian (i.e. p = 0, b = 1, $\upsilon$ = 1). (c) Spectrum when proper diagonal boundary term is used (i.e. p = 0, b = 0, $\upsilon$ = 1). (d) Spectrum (red) when relaxation of coupling near the edge is included (i.e. p = 0, b = 0, $\upsilon$ = 1/2). For comparison, in panels (b-d), the black lines show the spectrum when there is no edge, and periodic boundary conditions are used *(i.e. p = 1, b = 1, $\upsilon$ = 1)*.

Since the honeycomb lattice is bipartite, it is a natural candidate to search for the topological Su-Schrieffer-Heeger (SSH) edge states. Indeed, the fermionic honeycomb lattices possess zero-energy edge states both in 2D and 3D settings. Naively, since edge states are analyzed on a single-particle level, one could expect that the edge states in magnetic lattices are an identical copy of the edge states in the corresponding fermionic lattices. Below, we demonstrate that this is not the case. We calculate the energy spectrum of a 2D magnetic lattice in a ribbon geometry and find a significant departure from the fermionic analogues.

In order to study edge states, let us consider a honeycomb ribbon of finite width with $N = 40$ atoms in the $y$ direction and being infinite in the $x$ direction as shown in Fig. 4. We do the Fourier transform in the $x$ direction and introduce the corresponding momentum $k_x$. We keep the real space representation in the $y$ direction and write the tight-binging N-by-N Hamiltonian as

$$H = JS \begin{pmatrix} 2\nu+b & -2\nu\cos k_x & & & -pb \\ -2\nu\cos k_x & 2\nu+1 & -1 & & \\ & -1 & 3 & -2\cos k_x & \\ & & -2\cos k_x & \ddots & \vdots \\ -pb & & & \ldots & 2\nu+b \end{pmatrix} \quad (36)$$

acting on the N-component wavefunction $\psi_j$, $j=1,\ldots,N$, where $j$ labels atoms in the y-direction. The Hamiltonian describes an effective 1D dimerized lattice parametrically-dependent on momentum $k_x$. Note the structure of the Hamiltonian (36): The off-diagonal elements alternate between $1$ and $2\cos k_x$. This represents the dimerized nature of the Hamiltonian, i.e. if viewed in the y direction the vertical bonds alternate with tilted bonds. The non-zero diagonal elements in Eq. (36) originate from the interatomic coupling $S_i^z S_j^z$ according to the Holstein-Primakoff transformation (2). This is peculiar because the interatomic coupling between spins generates the on-site (diagonal) potential in the bosonic language. Therefore, the diagonal element corresponding to each site is equal to the sum over all bonds connecting to this site,

$$H_{ii} = S\Sigma_i J_{ij} \quad (37)$$

This guarantees that the spectrum of the magnon Hamiltonian is positive. We diagonalize the spectrum of Eq. (36) and plot it as function of $k_x$ in Fig. 4(b-d). For brevity of exposition, we introduced several parameters to be discussed: $p$ implements periodic boundary conditions, $b$ - an extra boundary potential, $\nu$ - relaxation of coupling at the boundary. First let us consider the periodic boundary condition in the y direction, which corresponds to connecting the dashed bonds (seen in panel a) to the other side of the sample. The periodic boundary conditions are implemented by the parameters $p=1, b=1, \nu=1$. We plot the corresponding spectrum with black lines in all panels (b-d). The spectrum is symmetric and has two Dirac degeneracies expected for the periodic boundary condition. Let us know eliminate the dashed bonds in panel (a), i.e. consider a ribbon. We set the parameter $b=0$, which corresponds to cutting the dashed bonds shown in panel (a), and plot the corresponding spectrum with red in Fig. 4(b). We observe that the two Dirac cones are connected by the topological dispersionless edge states, which was broadly discussed in the context of fermionic honeycomb lattices. However, setting $p=0$ violates the rule (37) for magnons. In order to capture the open boundary condition correctly, we set $b=0$, which manifestly eliminates the dashed bonds but also introduces the boundary potential on the edge atoms. The resulting spectrum is shown in Fig. 4(c). In contrast with panel (a), edge states are not visible because they are pushed into the bulk band by the boundary potential. Thus, the edge states known for fermionic honeycomb lattices do not survive for the case of magnons. We suggest that a relaxation at the surface can still generate edge states. Indeed, it is feasible that the coupling between the edge atoms, highlighted with blue in panel (a), is different. We implement this relaxation of the edge states at the surface with $\nu=1/2$ in Hamiltonian.

## D.2. Surface states in 3D.

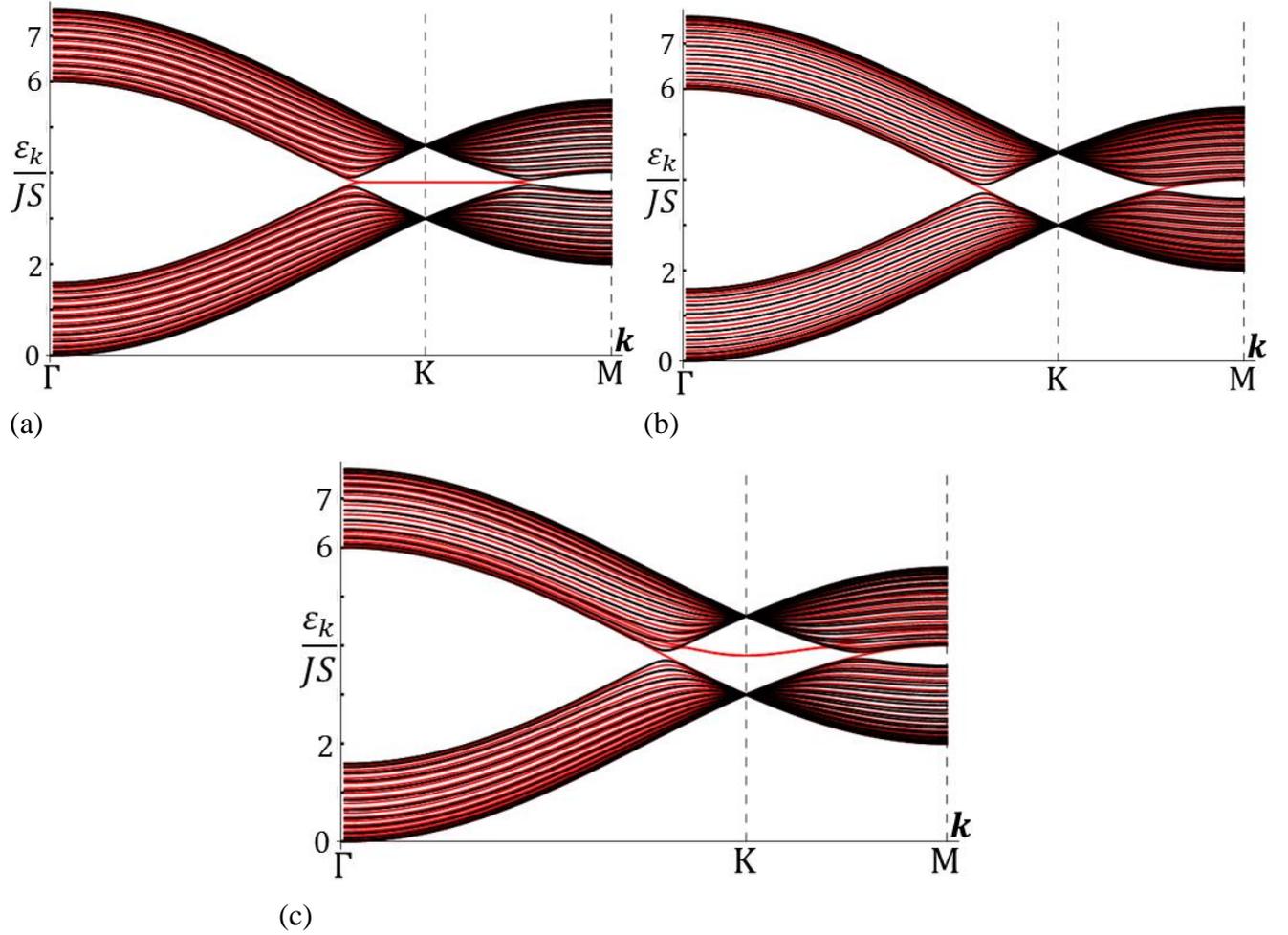

(a)

(b)

(c)

Figure 5. Magnon energy spectrum in a slab of 25 layers obtained numerically by diagonalizing Hamiltonian (38). (a) Red lines correspond to open boundary condition, i.e. parameters $p = 0$, $b = 1$, $v = 1$. This boundary condition is appropriate for fermions, but not magnons, (b) Red lines correspond to the correct boundary condition, i.e. parameters $p = 0$, $b = 0$, $v = 1$. The surface state is pushed into the bulk spectrum. (c) Red lines correspond to the case where relaxation of the interlayer coupling is introduced, i.e. $p = 0$, $b = 0$, $v = 0.5$. For reference, in all panels (a-b), the black lines correspond to the periodic boundary conditions, i.e. parameters $p = 1$, $b = 1$, $v = 1$ in Eq. (38).

In order to study the surface states, we follow the strategy taken in Sec. IIIA and study the magnon spectrum in a slab geometry. Then, the in-plane momentum $\mathbf{k} = (k_x, k_y)$ is a good quantum number, whereas we keep the real space representation in the $z$ direction

$$H = JS \begin{pmatrix} 3+b\gamma_z & -\gamma_\mathbf{k} & & & -pb\gamma_z \\ -\gamma_\mathbf{k}^* & 3+v\gamma_z & -v\gamma_z & & \\ & -v\gamma_z & 3+v\gamma_z & -\gamma_\mathbf{k} & \\ & & -\gamma_\mathbf{k}^* & \ddots & \vdots \\ -pb\gamma_z & & & \ldots & 3+b\gamma_z \end{pmatrix} \quad (38)$$

The parameter $p$ captures the periodic boundary condition, $b$ – the correct boundary term defined for magnons, $v$ implements the relaxation of the interlayer coupling z at the surface. The discussion identically follows Sec. III A. In panel (a), we show the spectrum of Hamiltonian (38) for open boundary conditions, i.e. $p = 0$. We observe the surface states flat band identical to the surface states for fermions. However, this boundary condition does not correctly describe surface states of magnons. The correct boundary condition, which corresponds to setting b = 0, sets an extra potential on the surface and pushes the edge states into the bulk spectrum (panel b). If we take into account the relaxation of the interlayer coupling $J_z$, by setting $v$ = 1/2, the edge states are "pulled" back in the gap (panel c).